\documentstyle[12pt,a4,epsf]{article} \textwidth 15.5 cm
\begin{document}
\bibliographystyle{plain} \input{epsf} {\Large\bf\noindent The Exact
Distribution of the Oscillation Period in the Underdamped One
Dimensional Sinai Model}
\vskip 2 truecm
\noindent David S. Dean and Satya N. Majumdar
\vskip 1 truecm
\noindent
IRSAMC, Laboratoire de Physique Quantique (UMR 5626 du CNRS),
Universit\'e Paul Sabatier, 118 route de Narbonne, 31062 Toulouse
Cedex 04, France.
\pagestyle{empty}
\vskip 1 truecm \noindent{\bf Abstract:} We consider the Newtonian
dynamics of a massive particle in a one dimemsional random potential
which is a Brownian motion in space. This is the zero temperature
nondamped Sinai model. As there is no dissipation the particle
oscillates between two turning points where its kinetic energy becomes
zero. The period of oscillation is a random variable fluctuating from
sample to sample of the random potential. We compute the probability
distribution of this period exactly and show that it has a power law
tail for large period, $P(T)\sim T^{-5/3}$ and an essential
singluarity $P(T)\sim \exp(-1/T)$ as $T\to 0$.  Our exact results are
confirmed by numerical simulations and also via a simple scaling
argument.

\vskip 1 truecm
\vskip 1 truecm
\noindent October 2001
\newpage
\pagenumbering{arabic}
\pagestyle{plain} Transport in random systems is an important and
common physical phenomenon. Even small amounts of disorder in physical
systems can dramatically modify their static and dynamical properties
and lead to interesting and complex new physics. In presence of
quenched, or time independent, disorder it is necessary to average
physical quantities over different realizations of the disorder. This
disorder averaging is often technically very difficult to carry out
and few exact results exist, even in one dimension. In this Letter we
study the zero temperature frictionless Newtonian dynamics of a
particle in a one dimensional random potential which is a Brownian
motion in space. Due to the absence of friction, the particle
oscillates back and forth between two turning points where the kinetic
energy becomes zero. The period of this oscillation is a random
variable, varying from sample to sample of the disorder. The main
result of this Letter is to present an exact result for the
distribution of this time period.

In general the dynamics of a particle in a one dimsional force field
can be represented by the Langevin equation
\begin{equation}
m{d^2 x\over dt^2} = -\gamma {dx\over dt} + F(x) + \eta(t),
\label{eqm}
\end{equation}
where $\gamma$ is the friction coefficient and $\eta(t)$ represents a
thermal noise with zero mean and a correlator $\langle
\eta(t)\eta(t')\rangle =2\gamma k_BT\delta(t-t')$, $T$ being the
temperature of the medium.  The force field $F(x)=-d\phi/dx$
originates from the background potential $\phi(x)$.  We will consider
the case when the potential $\phi(x)$ is a Brownian motion in
$x$. This means the force $F(x)$ is simply a Gaussian white noise in
space with zero mean and a correlator $\langle F(x)F(x')\rangle
=\delta(x-x')$.

The physics of this model is rather different in the two complimentary
limits namely the overdamped and the underdamped case. In the
overdamped limit where $m=0$, this model reduces to the celebrated
Sinai model\cite{Sinai}.  Since its introduction the Sinai model has
been studied extensively over the past two decades\cite{BG}.  It is
well known that the disorder drastically alters the diffusive
behaviour of the system and the mean squared displacement at any
nonzero temperature grows with time as $\overline{\langle
x_t^2\rangle} \sim \ln^4(t)$ for large $t$, where the angle brackets
here indicate the thermal average (over $\eta$) and the overline
indicates the disorder average (over $F$).

In the underdamped limit ($m>0$) this model represents, in the absence
of thermal fluctuations ($T=0$), the deterministic motion of a massive
particle in a random medium. Recently there has been a lot of interest
in understanding the dynamics of manifolds in a random
medium\cite{Fisher} where one neglects the thermal fluctuations as a
first approximation. The deterministic limit $\eta=0$ of
Eq. (\ref{eqm}) with $m>0$ then represents a toy model where one
considers the motion of a zero dimensional manifold.  For nonzero
friction $\gamma>0$, the particle will eventually stop at some
distance from the starting point since there is no thermal noise.
Recently Jespersen and Fogedby studied the distribution of this
stopping distance numerically and via simple scaling
analysis\cite{fogedby}. In the limit of vanishing friction $\gamma=0$,
this model was earlier studied by Stepanow and Schultz\cite{stepanow}
who showed that the particle resumes normal diffusion in dimensions
$d>1$ provided it starts with a nonzero initial velocity
$v_0$. However, in one dimension with $\gamma=0$, the particle does
not diffuse but oscillates between two turning points as mentioned
before. In this Letter we focus on this limiting case in one dimension
and calculate the distribution of the oscillation period exactly.

We set $m = 1$ for convenience and study simply the equation
\begin{equation}
{d^2 x\over dt^2} = F(x),
\label{eqm1}
\end{equation}
where $F(x)=-d\phi/dx$ is a white noise with zero mean and a
correlator $\langle F(x)F(x')\rangle=\delta(x-x')$. The potential
$\phi(x)=\phi_0+\int_0^x F(x')dx'$ is then a Brownian motion in $x$
which starts at the value $\phi(0)=\phi_0$ at $x=0$.  Without loss of
generality we set $\phi_0=0$.  We assume that the particle starts at
$t=0$ at $x=0$ with an initial velocity $v>0$.  The particle will move
to the right until it reaches a turning point $x_c$ where $\phi(x_c) =
v^2/2$ and then it will move back down to the point $0$ where its
velocity will be $-v$ (see Fig. 1). We define by $T$ the time taken to
arrive at the point $x_c$ starting from $x=0$. The particle will then
move to the left till it similarly reaches the left turning point
$x'_c$, the time taken to do this is $T'$. The total period of
oscillation is then given by $T_{osc} = 2(T+T')$. Since $\phi(x)$ is a
Brownian motion in $x$, it follows from its Markov property that
$\phi(x)$ for $x>0$ and $x<0$ are completely independent of each
other. Thus $T$ and $T'$ are also independent and has the identical
distribution which we compute below. The distribution of $T_{osc}$
then follows simply as the convolution of the distribution of $2T$
with that of $2T'$.

It is easy to express $T$ explicitly in terms of the potential
$\phi(x)$,
\begin{equation}
T = \int_0^{x_c} {dx\over \sqrt{v^2 - 2 \phi(x)}},
\label{eqintT}
\end{equation}
where we have used $\phi_0=0$.  Since $\phi(x)$ is random, the period
$T$ is also a random variable whose distribution we denote by
$P(T)$. Even though in the physical problem the turning time $T$ is
given by the formula in Eq. (\ref{eqintT}) with $\phi_0=0$, it is
useful to consider the random variable $T$ in Eq. (\ref{eqintT}) as a
function of an arbitrary $\phi_0$ and compute the distribution
$P(T,\phi_0)$ that depends on $\phi_0$. Eventually the distribution of
the physical time $T$ is given by $P(T)=P(T,0)$.  We define the
Laplace transform
\begin{equation}
Q(\mu, \phi_0) = \int_0^\infty P(T,\phi_0) \exp(-\mu T) \ dT.
\label{lt}
\end{equation}
Substituting the explicit form of $T$ from Eq. (\ref{eqintT}) in the
Laplace transform one finds
\begin{equation}
Q(\mu, \phi_0) = E_{\phi_0}\left[ \exp\left( -\mu \int_0^{x_c}
V(\phi(x)) dx\right)\right]
\label{lt1}
\end{equation}
where $E_{\phi_0}$ denotes the expectation value over Brownian paths
starting at $\phi_0$ at {\em time } $x = 0$ and
\begin{equation}
V(\phi(x)) = {1\over \sqrt{v^2 - 2 \phi(x)}}.
\label{vphi}
\end{equation}

There is a nice probabilistic interpretation of $Q(\mu, \phi_0)$ in
Eq. (\ref{lt1}).  Consider a Brownian trajectory $\{\phi(x) \ : 0\leq
x \leq \infty\}$ starting at $\phi_0$ at $x=0$. If the particle is
killed with probability $\mu V(\phi(x)) dx $ in the {\em time}
interval $[x,x+dx]$, then $Q(\mu, \phi_0)$ is simply the probability
that the particle is not killed before reaching the level $\phi =
{v^2\over 2}$.  In this interpretation $x_c = H_{{v^2\over 2}}$, where
$H_a$ is the first hitting time at level $a$, i.e., $H_a =
inf_{x>0}\{x\ : \phi(x) = a\}$.

To calculate $Q(\mu, \phi_0)$ we employ the standard backward Fokker
Planck technique \cite{gardiner,oksendal} in the following way. One
first evolves the Brownian motion over an infinitesimal {\it {time}}
interval $dx$ from the initial {\it {time}} $x=0$. The Brownian path
starts at $\phi_0$ at $x=0$. In {\it {time}} $dx$ it evolves to a new
position $\phi_0+d\phi_0$.  Thus the initial position of the
trajectory that starts at {\it time} $dx$ is $\phi_0+d\phi_0$.  As a
result of this infinitesimal change the function $Q(\mu,\phi_0)$
evolves as
\begin{equation}
Q(\mu, \phi_0)= \left(1 - \mu V(\phi_0)dx \right) E_{d\phi_0}\left[
Q(\mu, \phi_0 + d\phi_0) \right].
\label{evolve}
\end{equation}
The Eq. (\ref{evolve}) follows from the fact that the particle
survives in the interval $[0,dx]$ with probability $\left(1-\mu
V(\phi_0)dx\right)$ and then restarts its trajectory from the point
$\phi_0+d\phi_0$ at {\it time} $dx$.  Of course the variable $d\phi_0$
is random and one needs to sum over all histories of evolution in the
interval $[0,dx]$ which is denoted by the expectation $E_{d\phi_0}$ in
Eq. (\ref{evolve}).  Using the Ito prescription we average over
$d\phi_0$ to first order in $dx$ and then equating the terms of order
$dx$ yields the time independent Schrodinger equation
\begin{equation}
{1\over 2} {d^2 \over d\phi_0^2}Q(\mu,\phi_0) -\mu V(\phi_0)
Q(\mu,\phi_0) = 0,
\label{eqP1}
\end{equation}
where $V(\phi)$ is given by Eq. (\ref{vphi}). The function $Q$
satisfies the following boundary conditions: (i) $Q(\mu,v^2/2) =
1$. This follows from the fact that if $\phi_0=v^2/2$, then by
definition the hitting time $H_{{v^2\over 2}} = 0$ and subsequently
$x_c=0$ in Eq. (\ref{lt1}). (ii) $Q(\mu, -\infty) = 0$, as if $\phi(0)
\to -\infty$, the hitting time $H_{{v^2\over 2}} \to \infty$ implying
$x_c\to \infty$. Subsequently the integral in Eq. (\ref{lt1}) diverges
since $V(\phi(x))>0$ even when $x\to \infty$ and one obtains this
boundary condition.

Making the change of variables $y = \sqrt{v^2 - 2 \phi_0}$ in
Eq. (\ref{eqP1}) we get
\begin{equation}
{d^2 Q\over dy^2} -{1\over y} {dQ\over dy} - 2\mu y Q = 0.
\label{Qy1}
\end{equation}
The general solution to this equation is \cite{grad}
\begin{equation}
Q(y) = A y K_{2/3}\left[ \sqrt{{8 \mu y^3\over 9}}\right] + B y
I_{2/3}\left[ \sqrt{{8 \mu y^3\over 9}}\right],
\label{Qy2}
\end{equation}
where $A$ and $B$ are constants and $K_{\nu}(z)$ and $I_{\nu}(z)$ are
modified Bessel functions of index $\nu$. As $\phi_0\to -\infty$,
$y\to \infty$. In this limit the boundary condition (ii) implies the
constant $B=0$ since $I_{\nu}(z)$ diverges for large argument. The
boundary condition (i) implies that as $y\to 0$, $Q(y)\to 1$ which
fixes the constant $A=2^{4/3} \mu^{1/3}/{\Gamma({2/3}) 3^{2/3}}$.
Changing back to the $\phi_0$ variable, we then have the exact
expression for $Q(\mu,\phi_0)$
\begin{equation}
Q(\mu,\phi_0) = {2^{4\over 3} \mu^{1\over 3}\over \Gamma({2\over 3})
3^{2\over 3}} \sqrt{v^2-2\phi_0} K_{2/3}\left[ \sqrt{{8 \mu
(v^2-2\phi_0)^{3/2}\over 9}}\right].
\label{Qmu1}
\end{equation}
The distribution $P(T,\phi_0)$ is then obtained by inverting the
Laplace transform in Eq. (\ref{Qmu1}) which fortunately can be done
exactly \cite{grad}. Setting $\phi_0=0$ we finally obtain
\begin{equation}
P(T) = {2^{2\over 3} v^2 \over 3^{4\over 3} \Gamma({2\over 3})}
T^{-{5\over 3}} \exp\left[-{2 v^3 \over 9 T}\right].
\label{PT1}
\end{equation}
One can easily check that this distribution is normalized. In terms of
the scaled variable $\tau={9T\over 2 v^3}$ we find that the
probability density $p(\tau)=P(T)dT/d\tau$ of $\tau$ becomes
independent of $v$ and is given by the simple expression
\begin{equation}
p(\tau) = {1\over \Gamma({2\over 3})} {\tau}^{-{5\over
3}}\exp\left(-{1\over \tau}\right),
\label{eqtau}
\end{equation}
which is valid for all $t\ge 0$. In particular for large $\tau$,
$p(\tau)\sim {\tau}^{-\alpha}$ decays as a power law with the exponent
$\alpha = {5\over 3}$. Thus none of the integer moments of the
distribution exists and there are periods of all sizes extending upto
infinity.  The distribution $P(T)$ has a unique maximum at $T=T_m=
{2\over 15}v^3$ indicating the existence of a most probable turning
time $T_m$.

The exponent $\alpha=5/3$ can also be derived by an intuitive scaling
argument.  >From the energy conservation it follows that
$\phi(x_c)-\phi(x_0)=(dx_0/dt)^2/2$ where $x_c$ is the turning point,
$x_0$ is any other reference point and $dx_0/dt$ is the velocity at
$x_0$. For large $|x_c-x_0|$, clearly $\phi(x_c)-\phi(x_0)$ typically
scales as $\sim \sqrt{|x_c-x_0|}$ since $\phi(x)$ is a Brownian
motion. This indicates $dx_0/dt\sim |x_c-x_0|^{1/4}$ and hence
$|x_c-x_0|\sim t^{4/3}$ for large $t$.  The distribution of $x_c$
behaves as $\rho(x_c)\sim x_c^{-3/2}$ for large $x_c$ since $x_c$ is
simply the first passage {\it time} to the level $v^2/2$ of a Brownian
walker starting at the origin. It then follows that the distribution
of the turing time $t\sim |x_c-x_0|^{3/4}$ has a power law tail
$t^{-\alpha}$ with $\alpha=5/3$. In fact this scaling argument can be
generalized to an arbitrary random potential with a surface roughness
exponent $\chi$, such that $\phi(x)\sim x^{\chi}$ ($\chi=1/2$ for the
Brownian case). The distribution of the first passage {\it time} in
general has a power law decay, $\rho(x_c)\sim x_c^{-1-\theta_s}$ where
$\theta_s$ is the spatial persistence exponent\cite{MB}. Following the
same argument as in the Brownian case, one finds that the distribution
of turning time has a power law tail with an exponent
$\alpha=1+2\theta_s/(2-\chi)$.  For the Brownian case where $\chi=1/2$
and $\theta_s=1/2$, one recovers $\alpha=5/3$.

The distribution (\ref{eqtau}) has a rather remarkable link with the
original (over damped) Sinai model in the interval $[0,\infty]$, in
presence of an external force $F_{ext} = -f_0$ in the interval
$[0,\infty]$. Here the partition function in the canonical ensemble is
given by

\begin{equation}
Z = \int_0^\infty \exp\left[-\beta(\phi(x) + f_0 x)\right]dx
\end{equation}
where $\beta$ is the inverse temperature.  For $f_0 > 0$ the
distribution $\rho(Z)$ of $Z$ exists, can be computed
exactly\cite{comtet} and is given by
\begin{equation}
\rho(Z) = {\beta^2 \over 2\Gamma({2f_0\over \beta})}\left( 2\over
\beta^2 Z\right)^{1 + {2f_0\over \beta}} \exp\left(- {2\over \beta^2
Z}\right).
\end{equation}
Hence, when $f_0 = \beta/3 $ the partition function $Z$ (up to a
constant multiplicative factor) for the statics of the Sinai model in
the presence of a constant force $-f_0$ (towards the origin) has the
same distribution as the turning time $T$ in our model.

We have also verified our exact results via the numerical integration
of Eq. (\ref{eqintT}). We chose a discretization of the integral as
follows

\begin{eqnarray}
T(i+1) &=& T(i) + {\Delta x \over\sqrt{v^2 - 2 \phi(i)}} \\ \phi(i+1)
&=& \phi(i) + \sigma_i \sqrt{\Delta x}
\end{eqnarray}
Here the $\sigma_i$ are independent gaussian variables of mean zero
and variance one. The Brownian limit is clearly approached when
$\Delta x \to 0$.  This system is integrated up till the point $i$
where $\phi(i+1) > {v^2\over 2}$ and the value of $T(i)$
recorded. This is done for $5 \times 10^5$ realizations to construct
the histogram of the $T$.  The fact that $T$ has a broad distribution
implies that the number of integration steps required for a given
sample can become very large. Therefore a temporal cutoff $T^*$ is
placed on the values of $T$ measured. This means we discard those
samples where $T$ exceeds a large fixed value $T^*$.  Thus the
numerical histogram, once normalized, has the distribution
\begin{equation}
P(T,T^*) = {P(T) \theta(T^*-T)\over \int_0^{T^*} P(S) dS}.
\label{eqcutoff}
\end{equation}
In the limit when $T^*$ is large, $P(T,T^*)\to P(T)$. We first compute
$P(T,T^*)$ analytically from Eq. (\ref{eqcutoff}) using the exact form
of $P(T)$ from Eq. (\ref{PT1}). We then compare this analytical
result with the
numerically obtained $P(T,T^*)$ in Fig. (\ref{fignum}) using the
scaled variable $\tau=9T/2v^3$ with the choice $v=1$ and $T^*=5000$.
The agreement is evidently excellent.

\eject
\begin{figure}
\epsfxsize=1.0\hsize
\epsfbox{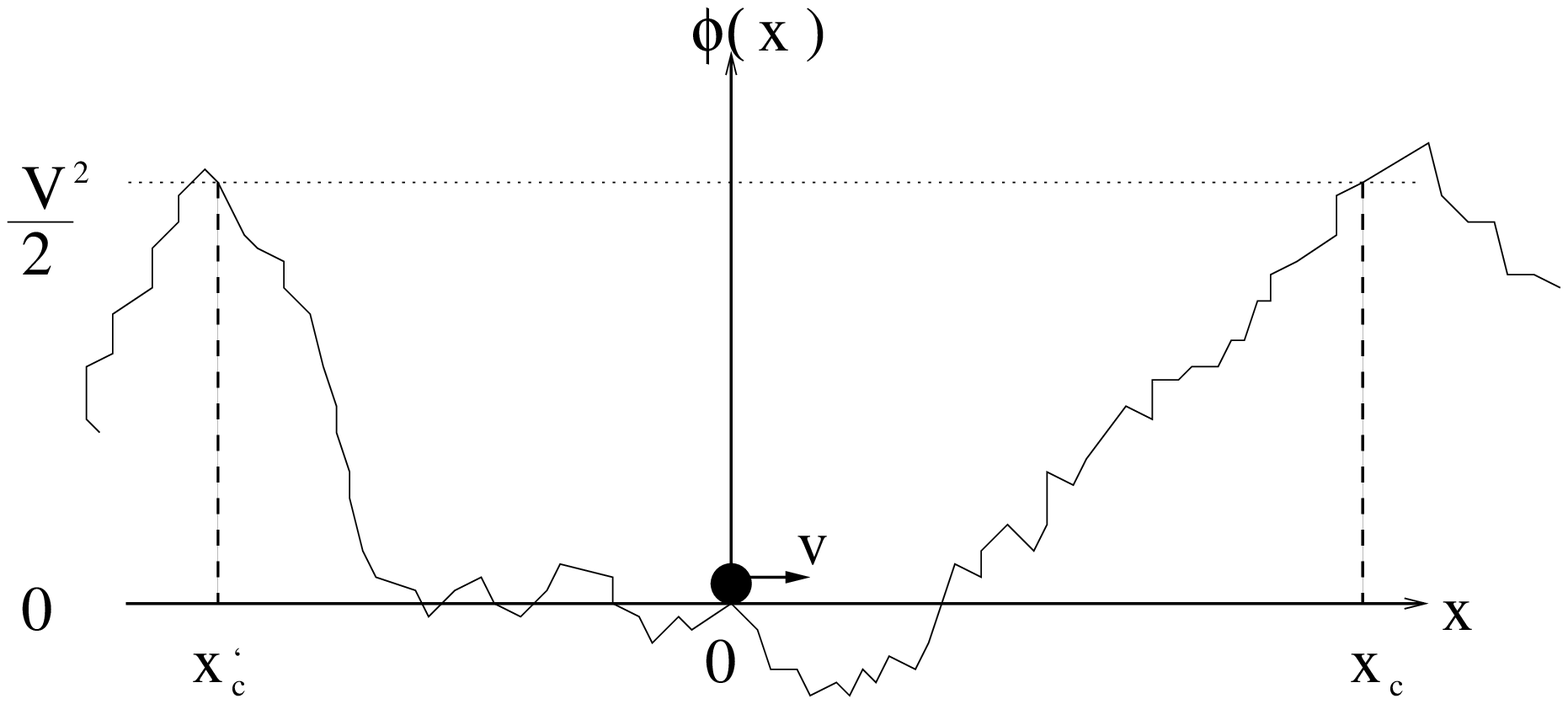}
\caption{ Particle in a Brownian potential with initial velocity $v$. The
right/left turning points are shown as $x_c$ and and $x'_c$ respectively
where the potential first becomes equal to $\phi = {v^2\over 2}$.}
\label{figpot}
\end{figure}
\eject
\begin{figure}
\epsfxsize=1.0\hsize
\epsfbox{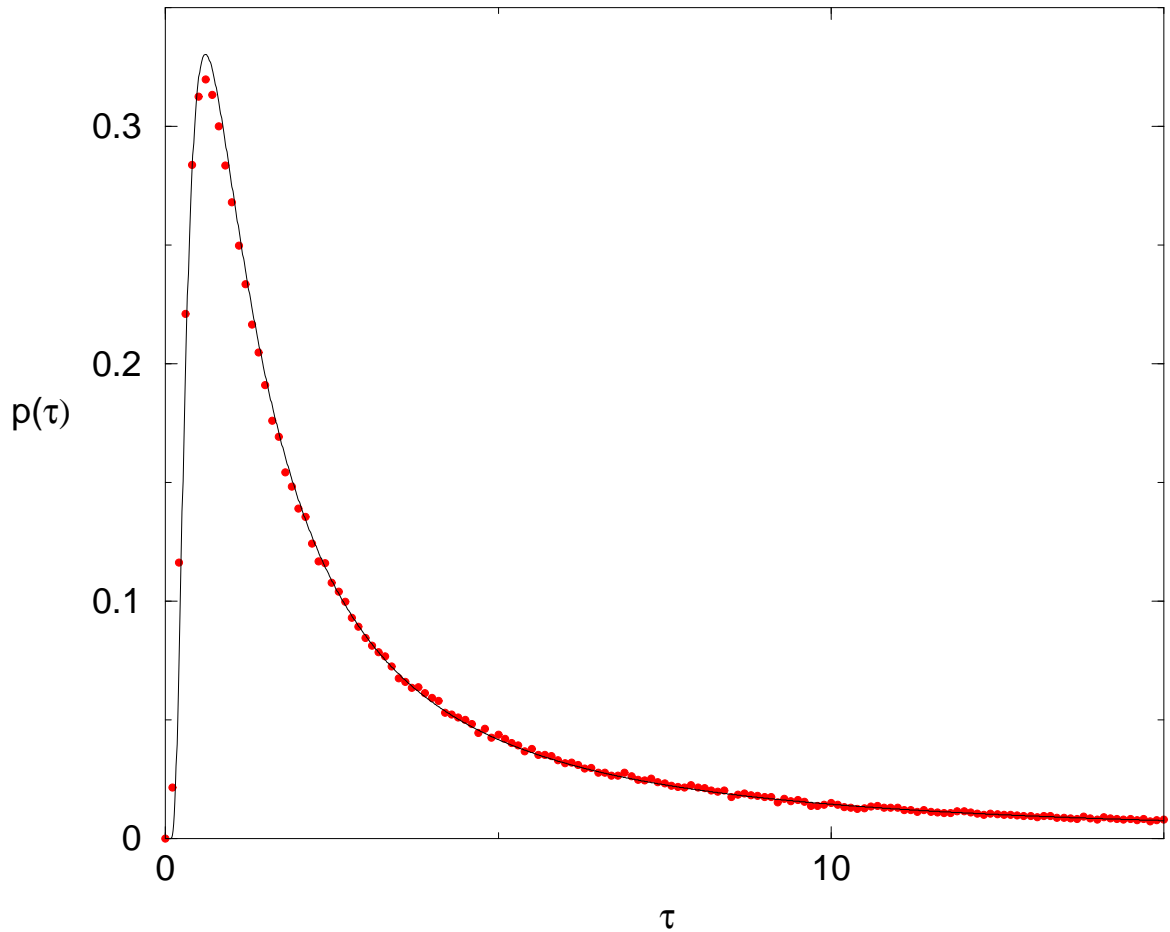}
\caption{Histogram of $\tau$ (circles) generated from $5\times 10^5$ samples,
with initial velocity $v=1$ and with temporal cutoff $T^* = 5000$, the bin size
is $0.1$. Shown also is the analytical prediction.}
\label{fignum}
\end{figure}

\end{document}